\documentclass{emulateapj}

\shorttitle{Thermal emission} \shortauthors{Pe'er \& Ryde}

\newcommand{\cm}{\rm{\, cm}}

\newcommand{\keV}{\rm{\, keV }}
\newcommand{\MeV}{\rm{\, MeV }}

\newcommand{\llu}{\rm{\, erg \, s^{-1}}}
\newcommand{\beq}{\begin{equation}}
\newcommand{\eeq}{\end{equation}}
\newcommand{\ba}{\begin{array}}
\newcommand{\ea}{\end{array}}
\renewcommand{\L}{L_{52}}

\newcommand{\Gi}{\Gamma_{2}}

\newcommand{\D}{{\mathcal {D}}}

\def \etal{{\it et al.~}}

\begin{document}
\title{A Theory of Mulicolor Black Body Emission from Relativistically Expanding Plasmas} 

\author{Asaf Pe'er\altaffilmark{1}\altaffilmark{2} and Felix Ryde\altaffilmark{3}}

\altaffiltext{1}{Space Telescope Science Institute, 3700 San Martin
  Dr., Baltimore, Md, 21218; apeer@stsci.edu}
\altaffiltext{2}{Giacconi Fellow}

\altaffiltext{3}{Department of Physics, Royal Institute of Technology,
AlbaNova, SE-106 91 Stockholm, Sweden}

\begin{abstract}

  We consider the emission of photons from the inner parts of a
  relativistically expanding plasma outflow, characterized by a constant
  Lorentz factor, $\Gamma$. Photons that are injected in regions of
  high optical depth are advected with the flow until they escape at
  the photosphere. Due to multiple scattering below the photosphere,
  the locally emerging comoving photon distribution is
  thermal. However, as an observer sees simultaneously photons emitted
  from different angles, hence with different Doppler boosting, the
  observed spectrum is a multi-color black-body. We calculate
  here the properties of the observed spectrum at different observed
  times.  Due to the strong dependence of the photospheric radius on
  the angle to the line of sight, for parameters characterizing
  gamma-ray bursts (GRBs) thermal photons are seen up to tens of
  seconds following the termination of the inner engine.  At late
  times, following the inner engine termination, both the number
  flux and energy flux of the thermal spectrum decay as $F \propto
  t^{-2}$. At these times, the multicolor black body emission results
  in a power law at low energies (below the thermal peak), with power
  law index $F_\nu \propto \nu^{0}$.  This result is remarkably
  similar to the average value of the low energy spectral slope index
  (``$\alpha$'') seen in fitting the spectra of large GRB sample.
\end{abstract}

\keywords{gamma rays:theory---plasmas---radiation
  mechanisms:thermal---radiative
  transfer---scattering---X-rays:bursts}

\section{Introduction}
\label{sec:intro}

Relativistic outflows, often in the forms of jets, are a common
phenomena in many astronomical objects, such as microquasars
\citep{MiRo94, HR95}, active galactic nuclei \citep[AGNs;
][]{LB85,GDSW06} and gamma-ray bursts \citep[GRBs; ][]{Pac86,Good86}.
In many of these objects, the density at the base of the flow is
sufficiently high, so that the optical depth to Thomson scattering by
the baryon-related electrons is much larger than unity. As a result,
if a source of photons exists deep enough in the flow, the emerging
spectrum is inevitably thermal or quasi-thermal (a Wien spectrum could
also emerge if the number of photons is conserved by the radiative
processes).  These photons escape the flow once they decouple from the
plasma, at the photosphere \citep[e.g.,][]{Pac90}.

As a result of the relativistic expansion of the source, even if the
emitted spectrum (in the plasma comoving frame) is purely thermal -
i.e., it is not modified by any additional, non-thermal radiative
process - still the observed spectrum is not expected to follow the
Planck distribution. This results from the aberration of light
in the expanding plasma, and is of pure geometrical nature. It is
therefore an inherent property of any relativistically expanding
photon emitting source.  The origin of this effect lies in the fact
that at any given instance, an observer sees simultaneously photons
that emerge the expanding plasma from a range of radii and
angles. Therefore, each (thermal) photon has its own comoving energy,
and is seen with a particular Doppler shift. The resulting, integrated
spectrum is non-thermal.

Analysis of this effect begins with the non-trivial shape of the
photosphere in a relativistically expanding plasma source. By
definition, the photosphere is a surface in space which fulfills the
following requirement: the optical depth to scattering a photon
originating from a point on this surface and reaching the observer is
equal to unity. Although it is mathematically a two-dimensional
surface in space, for spherically symmetric wind, as is considered
here, this surface is symmetric with respect to rotation around the
axis to the line of sight. It is therefore appropriate to refer to it
as the ``photospheric radius'', which is a function of the angle to
the line of sight, $r_{ph}=r_{ph}(\theta)$.

Calculation of the photospheric radius for the scenario of a steady,
spherically symmetric, relativistic wind characterized by a constant
Lorentz factor $\Gamma$ was first carried by \citet{ANP91} and later
extended by \citet{Peer08}.  For relativistic winds, $\Gamma \gg 1$,
and small angles to the line of sight, $\theta \ll 1$, the calculation
results in a simple, yet non-trivial dependence of the photospheric
radius on the angle to the line of sight: $r_{ph}(\theta) \propto
(\Gamma^{-2} + \theta^2/3)$, where the proportionality constant
depends on the properties of the outflow.  It is thus found that in
relativistic expanding wind, the photospheric radius has a strong
dependence on the angle to the line of sight, a fact which leads to
several non-trivial consequences: for example, it was shown by
\citet{Peer08}, that for parameters characterizing GRBs, photospheric
emission can be expected up to tens of seconds (albeit with a
decreasing flux; see further discussion below).

By definition, the photospheric radius provides only a first order
approximation to the last scattering position (=decoupling position)
of the thermal photons. This is because, in principle, photons can be
scattered at any point in space in which electrons exist. Therefore, a
full description of the last scattering position and scattering angle
can only be done in terms of probability density function
$P(r,\theta)$. As was shown by \citet{Peer08} and will be further
discussed here, use of this function is essential in calculating the
spectrum and flux. The shape of the observed spectrum, for example,
depends both on the individual Doppler shifts of the observed photons,
but also on the number of photons that undergo the particular Doppler
shift, an information that is only held in the probability density
function.

The probability density function can be considered as an extension of
the standard use of the photospheric radius. Instead of considering a
surface in space from which thermal photons emerge, one considers the
entire space, weighted by the finite probability of a photon to emerge
from an arbitrary radius $r$ and arbitrary angle $\theta$.  Using this
function, \citet{Peer08} calculated the expected temporal decay laws
of the observed thermal photon flux and average temperature, following
an abrupt termination of the inner engine.  It was shown there, that
at late times the thermal flux is expected to decay as $F \propto
t^{-2}$ and the average temperature decays as $<T> \propto
t^{-2/3}$. The agreement found between these theoretical predictions
and the late time decay laws of the peak energy and flux observed
during GRB prompt emission \citep{Ryde04, Ryde05, RP09}, is one of the
key motivations in studying the properties of thermal emission in the
context of the GRB prompt emission phase.

On a more general ground, thermal emission may be crucial in
understanding the nature of GRB prompt emission.  In spite of being
studied for nearly two decades now, the origin of the prompt emission
in GRBs is still puzzling. In recent years it became clear that
synchrotron emission, the leading emission model \citep{RM94,
  Tavani96, Cohen97, SP97, PSM99} cannot account for the steepness of
the low energy spectral slopes seen \citep{Crider97, Preece98,
  Preece02, GCG03} \citep[see, however][]{Bosnjak09}. This motivated
some alternative ideas, such as reprocessing through heated cloud
\citep{DB00}, jitter radiation \citep{Med00} or decaying magnetic
field \citep{PZ06}.  Moreover, in order to account for the high
efficiency of the prompt emission seen \citep{Zhang07, NFP09} using
the synchrotron model, a highly efficient energy dissipation is
required, which is difficult to be accounted for in the classical
internal shocks scenario \citep{KPS97, DM98, LGC99, GSW01}.
Contribution from thermal emission thus seems a natural way of
overcoming both these issues. First, photospheric emission is inherent
to the fireball model \citep{EL00, MR00, MRRZ02, DM02, RM05}. Second,
as it does not originate from any internal dissipation, contribution
from thermal photons reduces the efficiency requirement
\citep{RP09}. Finally, as will be shown here, photospheric emission is
able to produce low energy spectral slopes which are consistent with
those observed.

In spite of the success of the thermal emission model in reproducing
the late time decay of the peak energy and flux seen in many GRB's
\citep{RP09}, the simple version of the model suffers several
drawbacks. One issue that is often raised, is that the low energy
spectral slopes seen during the prompt phase of many GRBs are too
shallow to be accounted for by the Rayleigh-Jeans tail of the thermal
spectrum \citep[e.g.,][]{Bellm10}. Indeed, the Rayleigh-Jeans tail
implies a spectral slope $F_\nu \propto \nu^2$, while GRB observations
show that on the average, the low energy photon index in the ``Band''
function fits is $\alpha \simeq -1$ \citep[i.e., $F_\nu \propto
\nu^0$; see][]{Preece00, Kaneko06, Kaneko08}.

As discussed above, the observed spectrum resulting from a
photospheric emission in relativistically expanding plasma does not
necessarily need to be a pure black-body, but should, in general, be
modified.  Therefore, a pure blackbody spectrum will in many cases not
be able to fit the observed spectrum.  The main purpose of this paper
is to calculate the observed spectrum resulting from photospheric
emission in a scenario of a steady, relativistic outflow.  In fact, as
we show below, at late times following the termination of the inner
engine, the resulting spectrum is expected to be close to a power law
below the thermal peak, with power law index $F_\nu \propto \nu^0$. We
point out that this result is remarkably similar to the average value
of the low energy spectral slopes seen in large samples of GRBs
\citep{Kaneko06,Kaneko08}. We thus may obtain a natural explanation to
this observational result, in a model that considers emission from the
photosphere, once the full spatial scattering positions and scattering
angles are taken into account (of course, with several limitations
which are discussed below).

This paper is organized as follows. In \S\ref{sec:r_ph} we briefly
discuss the properties of the photosphere, and the characteristic time
scales up to which thermal emission is expected. In \S\ref{sec:prob}
we describe the construction of the probability density
function. These sections closely follow the treatment by
\citet{Peer08}, and are given here for completeness. We then calculate
in \S\ref{sec:analytic} the observed spectrum at late times, and show
that the energy spectrum can be approximated as $F_\nu(t) \propto \nu^0
t^{-2}$.  We compare the analytical predictions to the numerical
results in \S\ref{sec:numerics}. We then discuss the implications of
our results on the observed GRB prompt emission spectra in
\S\ref{sec:summary}.

\section{Basic considerations: photospheric radius and characteristic
  time scales in relativistically expanding plasma wind} 
\label{sec:r_ph}

Consider the ejection of a spherically symmetric plasma wind from a
progenitor characterized by constant mass loss rate $\dot M$, that
expands with time independent velocity $v = \beta c$. The ejection
begins at $t=0$ from radius $r=0$, thus at time $t$ the plasma outer
edge is at radius $r_{out}(t) = \beta c t$ from the center. However,
here we assume that the plasma wind occupies the entire space, i.e.,
$r_{out}(t) \rightarrow \infty$\footnote{This assumption has very
  little effect on the obtained results; see discussion in
  \citet{Peer08}.}.  For constant ${\dot M}$ and $\Gamma$, at
$r<r_{out}$ the comoving plasma density is given by $n'(r) = {\dot
  M}/(4\pi m_p v \Gamma r^2)$, where $\Gamma = (1-\beta^2)^{-1/2}$.
We further assume that emission of photons occurs deep inside the flow
where the optical depth is $\tau \gg 1$, as a result of unspecified
radiative processes.  The emitted photons are coupled to the flow
(e.g., via Compton scattering), and are assumed to thermalize before
escaping the plasma once the optical depth becomes low enough.

Under these assumptions, it was shown by \citet{Peer08}, that the
optical depth of a photon emitted at radius $r$, angle to the line
of sight $\theta$ and propagates toward the observer is 
\beq
\tau(r,\theta) = {R_d \over \pi r}\left[{\theta \over \sin(\theta)} -
  \beta \right] \simeq {R_d \over 2 \pi r} \left( {1\over \Gamma^2}
+{\theta^2 \over 3}\right).
\label{eq:tau1}
\eeq

 Here, 
\beq
R_d \equiv {{\dot M} \sigma_T \over 4 m_p \beta c},
\label{eq:rd}
\eeq
$m_p$ is the proton rest mass and $\sigma_T$ is Thomson cross section.
The last equality in equation \ref{eq:tau1} holds for $\Gamma \gg 1$
and small angle to the line of sight, $\theta \ll \pi/2$, which allows
the expansion $\sin(\theta) \simeq \theta - \theta^3/6$.

The photospheric radius is obtained by setting $\tau(r_{ph},\theta)
=1$,
\beq
r_{ph}(\theta) \simeq  {R_d \over 2 \pi} \left( {1\over \Gamma^2}
+{\theta^2 \over 3}\right).
\label{eq:r_ph}
\eeq
Equation \ref{eq:r_ph} implies that for small viewing angle, $\theta \ll
\Gamma^{-1}$ the photospheric radius is angle independent, $r_{ph}
\simeq R_d/2\pi \Gamma^2$, while for large angles $\theta \gg
\Gamma^{-1}$, the photospheric radius is $r_{ph}(\theta) \simeq R_d
\theta^2/6\pi$.

{\it Characteristic observed times.}  Below the photosphere, the
photons are coupled to the flow, therefore their effective propagation
velocity in the radial direction is similar to the outflow velocity,
$\approx \beta c$.  Assuming that a photon is emitted at $t=0$, $r=0$,
it decouples the plasma at time $t=r/\beta c$.\footnote{This result
  heavily relies on the assumption of constant outflow velocity below
  the photosphere. In GRBs as well as other astronomical objects,
  acceleration episode is expected, which changes the characteristic
  time scale of photon emergence. Nonetheless, we use this simplified
  assumption here, and further discuss it in \S\ref{sec:summary}.}
Consider photons that propagate towards the observer at angle to the
line of sight $\theta$. These photons are observed at a time delay
with respect to a hypothetical photon that was emitted at $t=0$, $r=0$
and did not suffer any time delay (``trigger'' photon), which is given
by
\beq
\Delta t^{ob.}(r,\theta) = {r \over \beta c}\times [1 - \beta \cos
(\theta)].
\label{eq:time}
\eeq

For relativistic outflows, $\Gamma \gg 1$,
thermal photons emitted from the photospheric radius $r_{ph}(\theta)$
on the line of sight ($\theta=0$), are seen at a very short time delay
with respect to the trigger photon, which is given by 
\beq 
\Delta t^{ob.}(r_{ph},\theta=0)\equiv t_N \simeq {R_d \over 4 \pi \Gamma^4
  \beta c} \simeq 10^{-2} \L \Gi^{-5} \, \rm{s}.
\label{eq:t_N}
\eeq
Here, ${\dot M} = L/\Gamma c^2$, and typical parameters characterizing
GRBs, $L = 10^{52} \L {\rm \, erg s^{-1}}$ and $\Gamma = 100 \Gi$ were
used.

On the other hand, due to the strong angular dependence of the
photospheric radius, thermal photons emitted from the photosphere with
high angles to the line of sight, $\theta \gg \Gamma^{-1}$ (and
$\theta \ll 1$) are observed at a much longer time delay,
\beq
t^{ob.}(\theta \gg \Gamma^{-1}) \simeq {R_d \over 3 \pi \beta
  c}\left({\theta^2 \over 2}\right)^2 \simeq 30 \, \L \Gi^{-1}
\theta_{-1}^4 {\rm \, s}, 
\label{eq:t_max}
\eeq
where $\theta = 0.1 \theta_{-1}$~rad.  The time scale derived on the right
hand side of equation \ref{eq:t_max} is based on the estimate of the
jet opening angle in GRB outflow, $\theta \leq \theta_j \simeq 0.1$~rad
\citep[e.g.,][]{BKF03}. Equation \ref{eq:t_max} therefore shows that in
a relativistically expanding wind with parameters that can
characterize GRBs, thermal emission is expected up to tens of seconds
following the decay of the inner engine.

\section{Use of probability density function in calculating the
  emission properties} 
\label{sec:prob}

Any attempt of describing the photospheric emission must consider the
fact that photons have a finite probability of being scattered at
every point in space in which electrons exist (i.e., in the entire
space, and not only on the photospheric surface). This led
\citet{Peer08} to introduce the concept of a probability density
function, as a mathematical tool which is needed in calculating the
observed properties of the photospheric emission, for example the
temporal decay of the temperature and flux. Since, as we show here,
this is a fundamental concept which is necessary in an analysis of the
properties of the photospheric emission, and in particular calculation
of the expected spectrum, we briefly repeat in this section the basic
definition and use of this function, before calculating the spectrum
in \S\ref{sec:analytic}.

The thermal photons are advected with the flow below the photosphere,
until the last scattering event (the decoupling) takes place, at some
radius $r$. For every radius $r$ there is an associated probability
that the last scattering event occurs at that particular radius (see
below).  During the last scattering event, a photon is scattered into
angle $\theta$. An underlying assumption is that photons that decouple
from the plasma at radius $r$ and scattered into angle $\theta$ are
observed at a delay given by equation \ref{eq:time}. This assumption
implies that: (a) the delay time of a photon is solely determined by
two parameters, the last scattering radius and scattering angle (for
constant outflow velocity); and (b) at any given instance, an observer
sees simultaneously photons emitted from a range of radii and angles,
all fulfilling the requirement set by equation \ref{eq:time}.

The $\theta$-dependence of the photospheric radius implies that the
probability of a photon to be scattered into angle $\theta$ depends on
the radius at which the scattering event takes place (or vice
versa). However, here we assume that the probabilities are
independent, i.e., $P(r,\theta) = P(r) \times P(\theta)$. This
assumption is made in order to simplify the calculation, and is tested
against the numerical results (see \S\ref{sec:numerics} below). While
clearly this approximation has only a limited validity, the results
obtained are in good agreement with the precise calculation done
numerically. We further discuss this approximation, as well as its
limitations in \S\ref{sec:summary} below.

Equation \ref{eq:tau1} implies that the optical depth to scattering
depends on the radius as $\tau(r) \propto r^{-1}$. This optical depth
is the integral over the probability of a photon propagating from
radius $r$ to $+\infty$ to be scattered, i.e., $\tau(r) =
\int_r^{\infty} (d\tau/dr) dr$, from which it is readily found that
$(d\tau/dr)|_r \propto r^{-2}$. As a photon propagates in the radial
direction from radius $r$ to $r + \delta r$, the optical depth in the
plasma changes by $\delta \tau = (d\tau/dr)|_r \delta r$. Therefore,
the probability of a photon to be scattered as it propagates from
radius $r$ to $r + \delta r$ is given by
\beq
P_{sc.} ( r .. r + \delta r) = 1 - e^{-\delta \tau} \approx \delta
\tau \propto {\delta r \over r^2}.
\label{eq:P_r1} 
\eeq    

For the last scattering event to take place at $r.. r+ \delta r$, it
is required that the photon does not undergo any additional scattering
before it reaches the observer. The probability that no additional
scattering occurs from radius $r$ to the observer is given by
$\exp(-\tau[r])$. The probability density function $P(r)$ for the last
scattering event to occur at radius $r$, is therefore written as
\beq
P(r) = {r_0 \over r^2} e^{-(r_0/r)}.
\label{eq:P_r}
\eeq 
The function $P(r)$ in equation \ref{eq:P_r} is normalized,
$\int_0^{\infty} P(r) dr = 1$.  Comparison to equation \ref{eq:tau1}
gives the proportionality constant, $r_0 \equiv r_{ph}(\theta=0) =
R_d/2\pi\Gamma^2$.

The probability of a photon to be scattered into angle $\theta$ is
calculated assuming isotropic scattering in the comoving frame, i.e.,
$d\sigma/d\Omega' = Const$.\footnote{This assumption neglects the
  dipole approximation, and is checked numerically to be valid.}  The
comoving spatial angle is $d\Omega' = \sin \theta' d\theta' d \phi'$,
and therefore the probability of a photon to be scattered to angle
$\theta'$ (in the comoving frame) is $dP/d\theta' \propto \sin
\theta'$. The proportionality constant is obtained by integrating over
the range $0 \leq \theta' \leq \pi$, and is equal to $1/2$. Thus, the
isotropic scattering approximation leads to $P(\theta') = (\sin
\theta')/2$.

Assuming that on the average, photons propagate in the radial
direction, by making Lorentz transformation to the observer frame one
obtains the probability of scattering into angle $\theta$ with respect
to the flow direction. This angle is equal to the observed angle to
the line of sight, and is given by
\beq
P(\theta) = P(\theta') {d\theta' \over d\cos\theta'} 
{d  \cos\theta' \over d \cos\theta} {d \cos\theta \over d\theta} = 
{\sin\theta \over 2 \Gamma^2 (1-\beta \cos\theta)^2}. 
\label{eq:P_theta}
\eeq
Defining $u \equiv 1-\beta \cos \theta $, equation \ref{eq:P_theta}
becomes 
\beq
P(u) = { 1 \over 2 \Gamma^2 \beta u^2}.
\label{eq:P_u}
\eeq 
Note that $1-\beta \leq u \leq 1+\beta$, and the function $P(u)$
in equation \ref{eq:P_u} is normalized, $\int_{1-\beta}^{1+\beta} P(u)
du = 1$.

\subsection {Spectrum and decay law of the thermal flux at late times}
\label{sec:analytic}

As long as the radiative processes that produce the thermal photons
deep inside the flow are active, the observed thermal radiation is
dominated by photons emitted on the line of sight towards the
observer. Once these radiative processes are terminated, the radiation
becomes dominated by photons emitted off axis and from larger radii,
which determine the late time behavior of the spectrum and flux.
Therefore, the limiting case of a $\delta$-function injection, both in
time and radius ($t=0$, $r=0$) is expected to closely describe the
late time behaviour of the thermal spectrum.  We calculate here the
observed spectrum and flux of the thermal emission at late times,
under these assumptions.

Denote by $T'(r)$ the photon comoving temperature, its observed
temperature is $T^{ob} = T'(r) \D$, where $\D = [\Gamma(1-\beta \cos
\theta)]^{-1} = (\Gamma u)^{-1}$ is the Doppler factor. The observed
photon temperature therefore depends on the viewing angle as well as
on the radius of decoupling. Below the photospheric radius, photons
lose their energy adiabatically, hence $T'(r) \propto r^{-2/3}$. It
was shown by \citet{Peer08}, that a similar decay law for the photon
temperature exists even if the energy density in the photon field is
much smaller than the energy density in the electrons (which can in
principle be non-relativistic in the comoving frame, hence have a
different temperature decay law), resulting from the spatial 3-d
expansion of the plasma. The adiabatic losses take place only as long
as the photons propagate at radii smaller than $\sim few \times r_0$.
While photons that propagate at high angles decouple the plasma at
much larger radii than $r_0$ (see eq. \ref{eq:r_ph}), above $few
\times r_0$, the number of scattering is small, and hence the photons
maintain their energy. As here we are interested in the late time
evolution of the spectrum and flux, where late time imply $t \gg t_N$,
we can safely assume that the comoving temperature of photons that
dominate the flux at late times is (on the average) constant, $T'(r) =
Const$.

Assume that $N_0$ photons are emitted instantaneously (a
$\delta$-function injection in time) at the center of the expanding
plasma. Each photon is advected with the flow, until the last
scattering event takes place at radius $r$ and into angle $\theta$,
after which it propagates freely.  The observed flux density (or
differential energy flux) from a source at luminosity distance $d_L$
is therefore given by
\beq
\ba{lcl}
F_{\nu}(t^{ob}) & \equiv  {d^2 F \over dt d\nu} =  &
{h N_0 \over 4 \pi d_L^2} \int P(r) dr \int P(u) du
T^{ob}(r,u)  \nonumber \\
& & \times \delta \left(t^{ob} = {r u \over \beta c} \right) \delta
\left(T^{ob} = {T'(r) \over \Gamma u} \right).
\ea
\label{eq:dE} 
\eeq 
Here, $F$ is the total fluence, and $h$ is Planck's constant; the
observed frequency $\nu$ corresponds  to the observed temperature, $\nu =
T^{ob}/h$. 

As discussed above, at late times, $t \gg t_N$, one can write $T^{ob}
= T'_0/\Gamma u$, where $T'_0$ is $r$-independent. Using $P(r)$ and
$P(u)$ from equations \ref{eq:P_r} and \ref{eq:P_u}, and the
identities $\delta(t^{ob} = ru/\beta c) = \delta (r=\beta c t^{ob}/u)
\times \beta c/u$ and $\delta (T^{ob} = T'_0/\Gamma u) = \delta (u =
T'_0 /\Gamma T^{ob}) \times \Gamma u^2 / T'_0$, equation \ref{eq:dE}
becomes
\beq
\ba{lcl}
F_{\nu}(t^{ob} \gg t_N) & = & {h N_0 \over 4 \pi d_L^2} \int {r_0 \over r^2} e^{-{r_0
    \over r}} dr \int {1 \over 2 \beta \Gamma^2 u^2} du
{T'_0 \over \Gamma u} \nonumber \\
& & \times {\beta c \over u} \delta \left(r = {\beta
    c t^{ob} \over u} \right) {\Gamma u^2 \over T'_0} \delta
\left(u = {T'_0 \over \Gamma T^{ob}} \right) \nonumber \\
& = & {h N_0 c \over 8 \pi d_L^2 \Gamma^2} \int {r_0 \over r^2} e^{-{r_0
    \over r}} dr \nonumber \\
& & \times \delta \left(r = {\beta c t^{ob} \Gamma T^{ob}\over
    T_0} \right) {\Gamma^2 {T^{ob}}^2 \over {T'_0}^2} \nonumber \\
& = & {h N_0 c \over 8 \pi d_L^2 r_0} \left({r_0 \over \Gamma \beta c t^{ob} }\right)^2
e^{-{r_0 T'_0 \over \Gamma \beta c t^{ob} T^{ob}}}.
\ea
\label{eq:f_nu1}
\eeq
This equation can further be simplified by using the definition of
$t_N$ from equation \ref{eq:t_N} and noting that at any given
instance, the maximum observed temperature is given by $T^{ob}_{\max}
= T'_0 / \Gamma u_{\min} = T'_0/\Gamma (1-\beta) \simeq 2 \Gamma
T'_0$. Using these in equation \ref{eq:f_nu1} leads to the final
form,
\beq
F_{\nu} (t^{ob} \gg t_N) = {h N_0 c \over 2 \pi d_L^2 r_0} \left({\Gamma t_N
    \over t^{ob} }\right)^2 e^{-{t_N \over t^{ob}}{\nu_{\max} \over \nu}},
\label{eq:f_nu}
\eeq
where $\nu_{\max} = T_{\max}^{ob}/h$. 

Equation \ref{eq:f_nu} is the key finding of this paper. It implies
that at late times, $t^{ob} \gg t_N$, for a wide frequency range
$\nu_{\max} (t_N/t^{ob}) < \nu < \nu_{\max}$,  the exponent is close to
unity, and therefore a flat energy spectrum
$F_\nu \propto \nu^0$ is expected below the thermal peak. This
spectrum results from a simultaneous observation of thermal photons
emitted from a large range of radii and angles to the line of
sight. We emphasis again that this is a purely geometrical effect, as
no additional radiative processes are considered. We further note that
equation \ref{eq:f_nu} provides, in addition, the decay law of the
energy flux at late times, $F_\nu(t^{ob} \gg t_N) \propto
{t^{ob}}^{-2}$.

\section{Numerical calculation of the flux and temperature decay at
  late times}
\label{sec:numerics}

The analytical calculations presented above were checked with a
numerical code. The code is a Monte-Carlo simulation, based on earlier
code developed for the study of photon propagation in relativistically
expanding plasma \citep{PW04, PMR06b}. This code is essentially
identical to the code used for the numerical calculations that appear in
\citet{Peer08}, and a description of it appears there. We give here
only a basic description of the code, for completeness, before
presenting the numerical results and a comparison to the analytical
approximation presented above.

The code is essentially a Monte-Carlo simulation of Compton scattering
between photons and electrons. The uniqueness of it lies in the fact
that it calculates the interactions during a relativistic,
three-dimension expansion of the plasma. As a result, the probability
of a photon to be scattered at any given instance, which is translated
to the distance traveled by a photon between two consecutive
scattering events, depends on the instantaneous radius and propagation
direction of the photon. In every interaction, the full Klein-Nishina
cross section is used in calculating the outgoing photon energy and
propagation direction. Since a scattering event is calculated in the
electron's rest frame, before and after every scattering, the photon
4-vector is being Lorentz transformed twice. First, into the (local)
bulk motion rest frame of the flow which assumes an expansion at
constant Lorentz factor $\Gamma$ in the radial direction. A second
Lorentz transformation is made into the electrons rest frame: in the
bulk frame, the electrons assume a random velocity direction, with
velocity drawn from a Maxwellian distribution with temperature
$T'_{el}(r) \propto r^{-2/3}$. The proportionality constant is
determined using parameters characterizing the prompt emission in
GRB's.

Initially, photons are injected into the plasma in a random position
on the surface of a sphere at radius $r_{inj} = R_d/(2\pi \Gamma^2
d)$. The depth $d$ is taken as $d=20$ in order to ensure that the
probability of a photon to escape without being scattered is smaller
than $\exp(-20)$, i.e., negligible\footnote{In fact, it was shown by
  \citet{Peer08} that the average number of scattering prior to photon
  escape is $\approx 2 d$.}. The initial photon propagation direction
is random, and its (local) comoving energy at the injection radius is
equal to the plasma comoving temperature at that radius.

\subsection{Numerical results}
\label{sec:numerical_results} 

\begin{figure}
\plotone{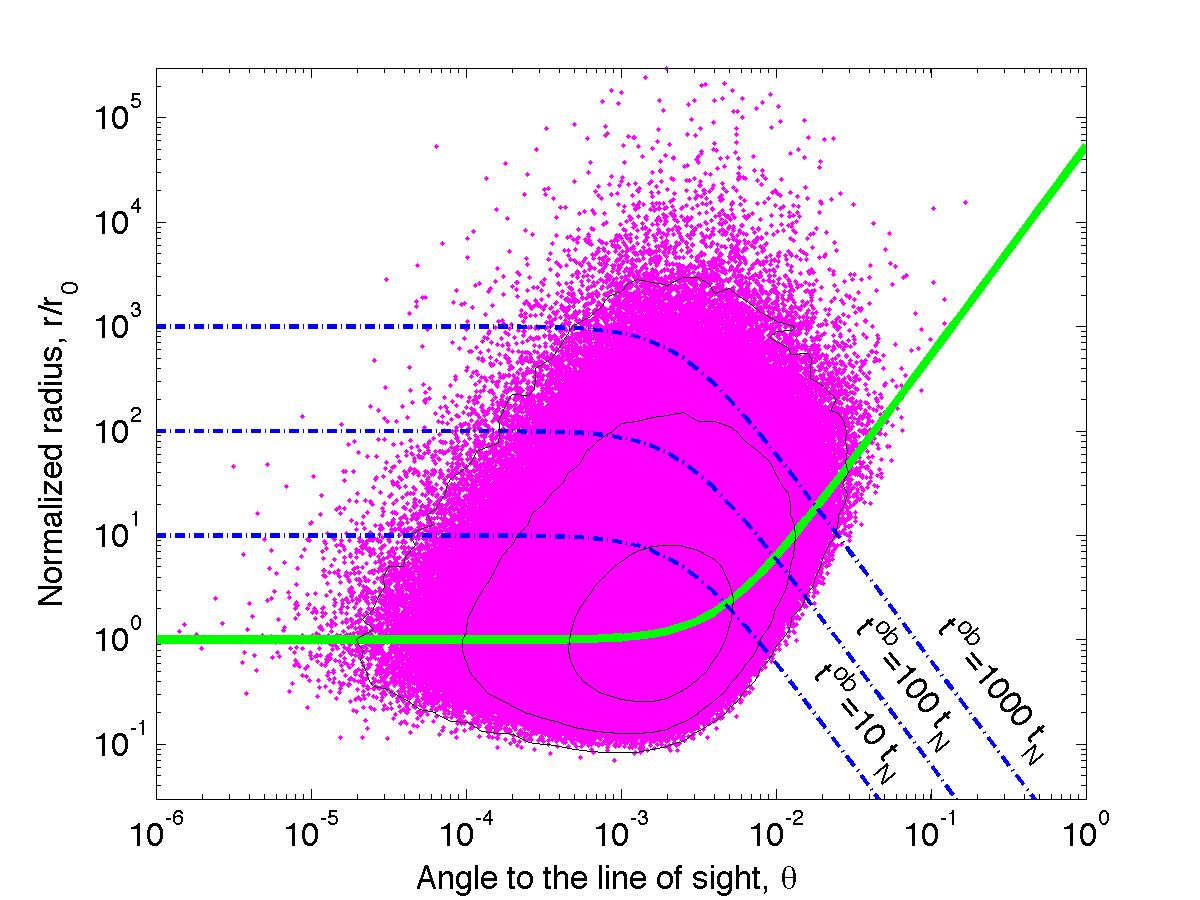}
\caption{Position of the last scattering event in $\theta$-$r$ plane
  for $10^{6.5}$ events. Parameters chosen in producing the plot are
  $L = 10^{52} \llu$ and $\Gamma=400$. The solid (green) line is the
  photospheric radius, calculated in equation \ref{eq:r_ph}. Clearly,
  the last scattering events occur over a wide range of radii and
  angles. The photospheric radius gives only a first order
  approximation to the position of these events. The contour lines are
  added to the plot in order to indicate the density of the emerging
  photons radii and angles. The blue (dashed) lines are equal arrival
  time contours, calculated in equation \ref{eq:t_N}. While at early
  times, $t \gtrsim t_N$ the emission is dominated by photons
  scattered at angles $\theta < \Gamma^{-1}$, at late times, photons
  emitted from large angles dominate the flux, a fact that gives rise
  to the flattening of the spectrum.}
\label{fig1}
\end{figure}

We show in figure \ref{fig1} the positions of the last scattering
events for $N=3\times 10^6$ photons in the $r-\theta$ plane. The three
contour lines (thin black) are added to the plot in order to help
demonstrating the probabilities of photons to be scattered at a given
radius and angle. The thick (green)
line, is the photospheric radius, as calculated in equation
\ref{eq:r_ph}. Clearly, the photospheric radius provides
only a first order approximation to the last scattering events
positions. It is obvious from the figure that photons decouple from
the plasma at a range of radii and angles, necessitating the use of
the probability density functions introduced in \S\ref{sec:prob}.  We
further added to the plot (dashed, blue lines), three equal arrival
times contours (see eq. \ref{eq:t_N}), for three different values of
the observed time. These contours demonstrate one of the key results:
while at early times, $t^{ob} \lesssim few \times t_N$, the emission
is dominated by photons emitted at angles smaller than $\Gamma^{-1}$,
at later times, most of the contribution to the emission is from
photons emitted from higher angles, $\theta > \Gamma^{-1}$. This is
the origin of the emerging power-law spectrum.

In preparing the plot, we used parameters that can characterized GRBs,
such as initial expansion radius $r_i = 10^8$~cm, luminosity $L =
10^{52} \llu$ and Lorentz factor $\Gamma=400$. In accordance to
equation \ref{eq:r_ph}, the photospheric radius is angle-independent
at angles smaller than $\Gamma^{-1} = 2.5 \times 10^{-3}$~rad, and is
$\propto \theta^2$ at higher angles.  We point out that in figure
\ref{fig1}, the radius of the last scattering event is normalized to
$r_0 = R_d/2\pi\Gamma^2$ ($r_0 = 9\times 10^{10} \cm$ for the
parameters chosen here). Therefore, results obtained for arbitrary
values of the free model parameters ($L$ and $\Gamma$) that
characterize astrophysical transients other than GRBs, such as AGNs or
microquasars are similar to the ones presented.

\begin{figure}
\plotone{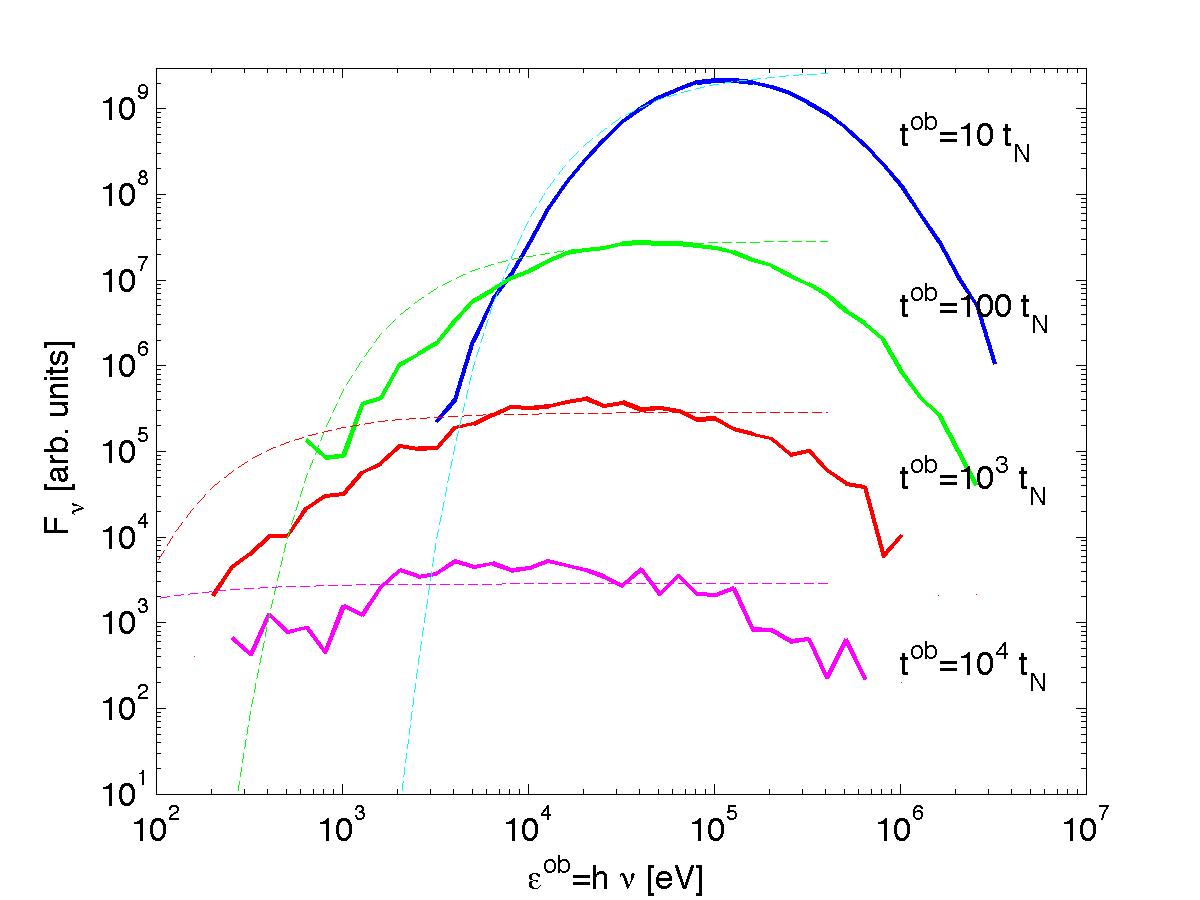}
\caption{Observed spectrum at different times, $t/t_N = 10, 10^2,
  10^3, 10^4$. The solid lines are the results of the numerical
  simulation, and the dashed (thin) lines are the analytical
  approximation derived in equation \ref{eq:f_nu}. For initial
  expansion radius $r_i = 10^8$~cm and other parameters same as in
  figure \ref{fig1}, as considered here, $T_{\max}^{ob} = 430 \keV$
  (see text for further details). At late times, $t \gg t_N$, the
  spectrum becomes flat, $F_\nu \propto \nu^0$ over a wide spectral
  range below $T_{\max}^{ob}$. This spectrum is very different than
  the Rayleigh-Jeans tail of a thermal spectrum.}
\label{fig2}
\end{figure}

In figure \ref{fig2} we present the resulting observed spectrum,
$F_\nu$ calculated at different times.  The solid lines show the full
numerical results at observed times $t^{ob}/t_N = 10, 10^2, 10^3,
10^4$. The dashed (thin) lines are the analytical approximation given
by equation \ref{eq:f_nu}. In calculating the maximum observed
temperature, $T_{\max}^{ob}$, we used the standard ``fireball'' model
to determine the temperature evolution below the photosphere
\citep[see, e.g.,][]{Piran05, Mes06}.  Thus, the plasma assumes to
accelerate between $r_i$ and the saturation radius $r_s = \Gamma r_i$,
and continues with constant outflow velocity at larger radii. For the
values of the parameters chosen, the plasma comoving temperature at
the photospheric radius $r_0$ is $T'(r_0) = (L/4\pi r_i^2 c a)^{1/4}
(r_0/r_s)^{-2/3} = 0.54 \L^{-5/12} r_{i,8}^{1/6} (\Gamma/400)^{5/3}
\keV$, and the maximum expected temperature is $T_{\max}^{ob} = 2
\Gamma T'(r_0) = 430 \keV$.

The numerical results are indeed in very good agreement with the
analytical approximation. We note that for the parameters chosen here,
which can characterize many GRBs, $t_N \approx 10^{-5}$~s, and
therefore the observed characteristic times of $\lesssim$~second (in a
single pulse) are translated into $t^{ob}/t_N \leq 10^4$.  At early
times, the exponential decay occurs at (relatively) high energies, and
thus a flat spectrum is not seen. However, at later times, the
spectrum becomes flat over a wide energy range, consistent with the
analytical approximations.

At intermediate times there is a slight discrepancy between the
numerical results and the analytical approximation, both at the high
and low energies.  This discrepancy demonstrates the limitation of the
delta-function approximation used in eq. \ref{eq:f_nu}. At high
energies, the numerical results show ``bending'' at early times, which
is not captured by the analytical approximation. This results from the
the $\delta$-function approximation in time used
(eq. \ref{eq:t_N}). The underlying assumption used in deriving
equation \ref{eq:t_N} is that the observed delay time is solely
determined by the radius and angle of the last scattering event. In
reality, of course, as the photons diffuse below the photosphere, the
time delay is determined by the full history of the photon propagation
below the photosphere. At high energies, the spectrum is dominated by
photons emitted with high Doppler shift, i.e., at small viewing
angles; as at late times there are relatively very few such photons,
dispersion in the last scattering position and angle leads to a high
energy decay.

At low energies, the discrepancy between the analytical prediction and
the numerical results is explained due to a combination of two
phenomena. First, the analytical approximation used does not consider
the coupling between the probability density functions $P(r)$ and
$P(\theta)$. As is clear from figure \ref{fig1}, for any given radius
$r$, the observed viewing angle $\theta$ is limited from above. A
second source of spread lies in the fact that the comoving photons
energy distribution also have an internal spread, (although, the
average comoving energy is constant above $few \times r_0$). As a
result of these two effects, the low energy spectrum is also slightly
bended with respect to the simple power law prediction of the
analytical approximation. For a full treatment of these effects, one
needs to obtain a full solution of the diffusion equation, which will
be presented elsewhere.

In spite of being only a first order approximation, it is clear from
figure \ref{fig2} that the analytical predictions are in very good
agreement with the numerical results of the observed spectrum.  At
intermediate times, the spectrum approaches a power law distribution
$F_\nu \propto \nu^0$ over a relatively wide energy band. Even at
very early times, $t \gtrsim t_N$, it is clear that the observed
spectrum deviates significantly from the classical Planck function.

\section{Summary and discussion}
\label{sec:summary}

In this manuscript, we have studied both analytically and numerically
the spectrum resulting from a photospheric emission in
relativistically expanding plasma. We showed that even for purely
thermal distribution of the comoving photon spectrum, aberration of
the light results in an observed spectrum at late times that is very
different from a Planck distribution. We showed that at late times, $t
\gg t_N$, the observed spectrum approaches a power law, with power law
index $F_\nu \propto \nu^0$ over a wide spectral range. This
result is remarkably similar to the observed low energy spectral index
of many GRBs \citep{Preece00, Kaneko06, Kaneko08}.

The origin of this non-trivial result lies in two facts: First, at
late times, the emission is dominated by photons emitted off-axis
(from angles $\theta \gg \Gamma^{-1}$, see figure \ref{fig1}). Due to
the strong dependence of the Doppler shift on the angle to the line of
sight, these photons are seen at much lower energies than photons
emitted on-axis.  Second, at any given instance, an observer sees
simultaneously photons that are emitted from a range of radii and
angles (see eq. \ref{eq:t_N}). As each photon has a finite probability
of being emitted from a given radius, and is seen at a particular
Doppler shift, the observed spectrum and flux can only be described in
terms of probability density functions (see \S\ref{sec:prob}). These
functions provide a mathematical tool to describe the probability of
photons to be emitted from radius $r$ and into angle $\theta$. Using
these functions, we calculated an analytical approximation to the
observed spectrum (eq. \ref{eq:f_nu}), which is the main result of
this work.  The analytical approximation was tested with a Monte-Carlo
simulation that tracks the evolution of thermal photons in
relativistically expanding plasma (\S\ref{sec:numerics}). The
simplified analytical calculations are found to be in very good
agreement with the accurate numerical results (see figure \ref{fig2}).

In spite of the success in reproducing the low energy spectral slopes at late times,
the theory is still not completed. For parameters characterizing GRBs,
the thermal peak ( $T_{\max}^{ob}$) naturally falls at the sub-MeV
range (see \S\ref{sec:numerical_results}), hence the observed peak can
naturally be explained as having a thermal origin. However, a
noticeable drawback of the theory as stated in this manuscript, is
that the same parameters lead to very short characteristic time scale,
$t_N \approx 10^{-5}$~s. Since above $t_N$ the flux decays rapidly,
$F_{\nu}(t) \propto t^{-2}$ (see eq. \ref{eq:f_nu}), one expects a
relatively weak thermal signal at $t \gg t_N$, which can still be very
short time.  We note though, that the calculation of $t_N$ in equation
\ref{eq:t_N} is based on the assumption of constant outflow
velocity. This assumption is too simplified: in order to reach high
Lorentz factor, the plasma needs to undergo an acceleration phase, and
so the average Lorentz factor below the photosphere is less than the
terminal Lorentz factor. As a result, we expect that in practice the
characteristic time scale relevant for thermal emission in GRBs is
longer than the one considered in equation \ref{eq:t_N}.

The exact delay time depends on several uncertain conditions. One is
the content of the fireball: for example, Poynting-flux dominated
fireball is expected to have slower acceleration than matter dominated
fireball \citep{Drenk02,DS02, GS05, GS06}. Another is the baryon load
\citep{Ioka10}, and in particular the baryon distribution along the
jet \citep{MLB07, LMB09, MNA10}: although in this manuscript we
considered a steady outflow, clearly the outflow in GRBs is
characterized by regions of higher and lower densities, and is thus
not steady (edge effects due to the finite opening angle may also play
a role at late times). 

An additional source of discrepancy between the theoretical
predictions developed in this paper and the observed spectrum, lies in
the fact that the theory here does not consider any additional,
non-thermal radiative processes. As shown here, photospheric emission
is capable of reproducing the peak energy and the low energy spectral
slope ($\alpha$ in the ``Band'' function) seen in GRBs (at late times). However,
photospheric emission is not capable of of producing high energy
photons (above $T_{\max}^{ob} \lesssim \MeV$), as are seen in some
GRBs by the LAT detector on board the {\it Fermi} satellite. The
inclusion of high energy, non-thermal photons, necessitates additional
radiative mechanisms, that must take place following dissipation
processes that occur above the photosphere \citep[e.g., in GRB080916C
analysis of the high energy emission imply Poynting dominated outflow;
see][]{ZP09}.  Additional radiative mechanisms, such as synchrotron
emission or Compton scattering, naturally produce a broad band energy
spectrum, and thus may contribute not only to the high energy spectrum
but to the low energy part (below the thermal peak) as well. The
overall observed spectra below the thermal peak is thus generally
expected to be hybrid- i.e., composed of both thermal and non-thermal
parts. The exact contribution of the non-thermal part may vary from
burst to burst, depending on the values of the free model parameters,
such as the radius of the photosphere, the dissipation radius, the
strength of the magnetic field, etc.

In principle, the inclusion of thermal photons contributes to the high
energy, non thermal part of the spectrum as well, as these photons
serve as seed photons for Compton scattering by the non-thermal
electrons and hot pairs \citep{RM05, PMR05, PMR06, LazBeg10,
  Bel10}. The exact contribution of the thermal photons depend on the
optical depth at the dissipation radius. If the dissipation radius is
close to the photosphere, the resulting spectrum has a complex shape,
that is very different than either a thermal spectrum or the optically
thin synchrotron - synchrotron self Compton (SSC) model predictions
\citep{PMR05, PMR06}. On the other hand, if the dissipation occurs at
large radii, the two components, thermal and non-thermal, can be
decoupled, a fact that can lead to a clear identification of the
thermal component, as is in the case of GRB090902B \citep{Ryde+10,
  Peer+10}.
 
One clear prediction of the results presented here, is that if a
thermal component contributed significantly to the observed spectrum,
than the low energy spectral slope ($\alpha$) is expected to vary with
time: at early times, when the inner engine is active, $\alpha$ is
expected to be steep (close to the Rayleigh-Jeans tail), while at
later times, $F_\nu \propto \nu^0$. This result is expected to be
correlated with a decay of the peak energy and the thermal flux, and
was possibly observed in several bursts \citep[e.g.,][]{Crider97,
  Ghirlanda+07, RP09, Page+09}. We note though that in many GRBs, the
data is not easily compared to the theoretical prediction, since the
outflows in GRBs are generally not smooth, but very
fluctuative. Therefore, photons originating from different emission
episodes (each characterized by its own mass loss ejection rate and
its own Lorentz factor) are often superimposed. Thus, in fact, thermal
photons may only be identified using detailed time resolved spectral
analysis of separate flares. This task was carried by \citet{RP09},
which were indeed able to identify the thermal component in a large
sample of bursts. Once done, this analysis method can be further used
to deduce the properties of the GRB outflow \citep{PRWMR07}.

In addition to the prompt emission phase in GRBs, thermal activity may
occur as part of the flaring activity observed in the early afterglow
phase of many GRBs \citep{Burrows05, Falcone07, Chincarini+10,
  Margutti+10}. The exact nature of these flares is currently not yet
clear. As it is plausible that the flares result from renewed emission
from the inner core, a renewed thermal emission may occur. The
analysis presented here may therefore apply in the study of the late
time flares as well.

While analysis of GRB prompt emission spectrum serves as a major
motivation to this work, we note that the analysis presented here is
general. With modified parameters values, the analysis apply to
emission from any astronomical transient, which is characterized by
relativistic outflow speeds and high density core. Thus, we expect the
analysis carried here to be valid for objects such as AGNs and
microquasars. This is valid because the exact nature of the radiative
process that produces the thermal photons is of no importance, as long
as the emission occurs deep enough in the flow, where the optical
depth is high, $\tau \gg 1$.

\acknowledgments 
We would also like to thank Mario Livio, {\u Z}eljka Bo{\u s}njak, Chryssa
Kouveliotou, Charles Dermer, Ralph A.M.J. Wijers, Peter
M\'esz\'aros, Martin Rees and Bing Zhang for many useful discussions
and comments. AP is supported by the Riccardo Giacconi Fellowship
award of the Space Telescope Science Institute.

\end{document}